\documentclass{aa}
\usepackage{graphicx}
\begin{document}
   \title{Is Messier 74 a barred spiral galaxy?}

   \subtitle{Near-infrared imaging of M74}

   \author{Marc S. Seigar$^1$}

   \offprints{Marc S. Seigar\\
              $^1$Guest investigator of the UK Astronomy Data Centre.}

   \institute{Joint Astronomy Centre, 660 N. A'ohoku Place,
              Hilo, HI 96720, USA\\
              \email{m.seigar@jach.hawaii.edu\\}
             }

   \date{Submitted 15-May-2002}

   \abstract{We have obtained ground-based {\em I}, {\em J} and {\em K} 
band images of the spiral
galaxy, Messier 74 (NGC 628). 
This galaxy has been shown to possess a circumnuclear ring
of star formation from both near-infrared spectroscopy of CO absorption and
sub-millimetre imaging of CO emission. Circumnuclear rings of star formation
are believed to exist only as a result of a bar potential. In this paper we 
show evidence for a weak oval distortion in the centre of M74. We use the 
results of Combes \& Gerin (1985) to suggest that this weak oval potential
is responsible for the circumnuclear ring of star formation observed in M74.
   \keywords{Galaxies: individual: M74 --
                Galaxies: fundamental parameters --
		Galaxies: spiral --
		Galaxies: structure
               }
   }

   \maketitle
%

\section{Introduction}

Circumnuclear rings of star formation have been shown to occur in barred 
spiral galaxies since the early 1980s (e.g. Benedict 1980) and they have been
studied in great detail in many galaxies since this pioneering work
(e.g. Knapen 1996; Knapen et al. 1999).
They are thought to be a
result of a funneling of material to the central regions of the galaxies
by a bar potential. Indeed, hydrodynamical simulations of galaxies have
shown that gaseous material is shocked at the leading edge of a bar and
diverted towards the centre of the galaxy (Roberts, Huntley \& van Albada
1979). The material can then accumulate at the inner Lindblad resonance (ILR)
until it reaches a critical density at which star formation can
be induced.

The spiral galaxy, Messier 74 (NGC 628), 
is classed as a non-barred spiral galaxy
(its Hubble classification is SAc - de Vaucouleurs et al. 1991). However, a 
circumnuclear ring of star
formation does exist in the central regions of M74. This has been observed
in $^{12}$CO $J=1-0$ 
sub-mm imaging (Wakker \& Adler 1995) and 2.3 $\mu$m CO absorption
spectroscopy (James \& Seigar 1999). Is it therefore possible that a bar--like
structure exists in the centre of M74, but is shrouded in dust? One way to
answer this question is to observe M74 at near-infrared wavelengths. 
The first demonstration that bars are more common in the near-infrared was
performed by Hackwell \& Schweizer(1983).
Since then, this method has
proved successful for uncovering bars in spiral galaxies in many cases (e.g.
Seigar \& James 1998; Eskridge et al. 2000). 
In this paper we present {\em I}, {\em J} and {\em K} band images
of M74, the longest wavelength images available, in order to uncover a bar in
its centre.

This paper is arranged as follows: section 2 describes the observations;
section 3 is a discussion of the results presented in this paper; in section
4 we summarise our main results.


\section{Observations}

In order to uncover a bar-like structure in the centre of M74, long 
wavelength images are needed. 
Ideally, near-infrared imaging would be the best 
tool for this kind of study. We have therefore retrieved archived 
{\em I}, {\em J} and {\em K} band images of M74.

   \begin{figure*}
   \begin{center}
   \includegraphics[width=16cm]{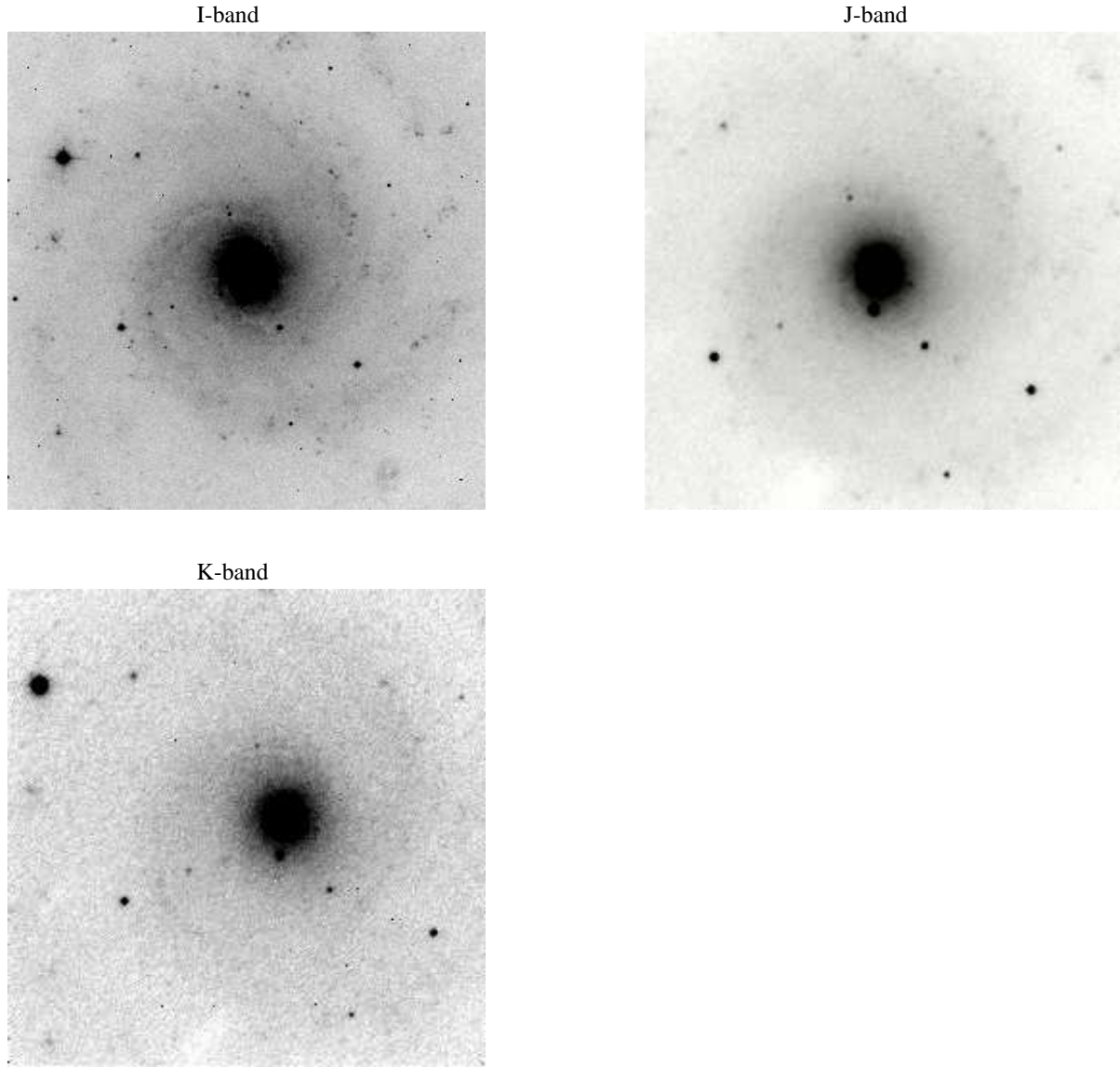}
   \caption{{\em I}, {\em J} and {\em K} band greyscale plots of Messier 74}
   \label{iimg}
   \end{center}
   \end{figure*}

We have made use of the Isaac Newton Group (ING) data archive to retrieve an 
{\em I} band image of M74. The data was observed on the Jacobus Kapteyn 
Telescope (JKT) on 9th November 1999. Two {\em I} band images of 300 seconds 
each were observed. These images have been flat-fielded (using twilight sky 
flats) and combined to produce a single image of effective exposure time 600 
seconds. We also used the NASA extragalactic database (NED) to retrieve {\em
J} and {\em K} band images of M74. These images were observed at the Calar
Alto 2.2m telescope using the MAGIC NICMOS3 instrument. For each waveband, M74
was observed for a total integration of 9 minutes. These images have been
flat fielded (using dome flats). The infrared images were originally 
presented in a paper by M\"ollenhoff \& Heidt (2001). 


\section{Discussion}

Figure 1 shows greyscale {\em I}, {\em J} and {\em K} band plots of 
M74 respectively. 
Figure 2 shows the 
equivalent contour plots. While it is not obvious in the greyscale plots, the 
{\em I} 
band contour plot clearly shows an oval distortion in the contours at a 
radius of 
approximately 100 pixels (or 33.0 arcsec for a pixel scale of 0.33 
arcsec/pixel) in the y-direction. The ellipticity of this contour is 
approximately 0.20$\pm$0.07. The {\em J} band and {\em K} 
band images both show this
elliptical contour as well, with ellipticities of 0.16$\pm$0.05 and 
0.17$\pm$0.05 respectively. The position angle of this contour is 
approximately 45$^{\circ}$. At a larger radius, of approximately 40--44
arcsec, the {\em J} and {\em K} 
band images also show an elliptical contour at a position
angle of -55$^{\circ}$. The ellipticities of this contour are 0.28$\pm$0.05
(in the {\em J} band image) and 0.33$\pm$0.05 (in the {\em K} band image).
As M74 is a face-on galaxy, the ellipticities calculated here are 
real and not an inclination effect.

   \begin{figure*}
   \begin{center}
   \includegraphics[width=16cm]{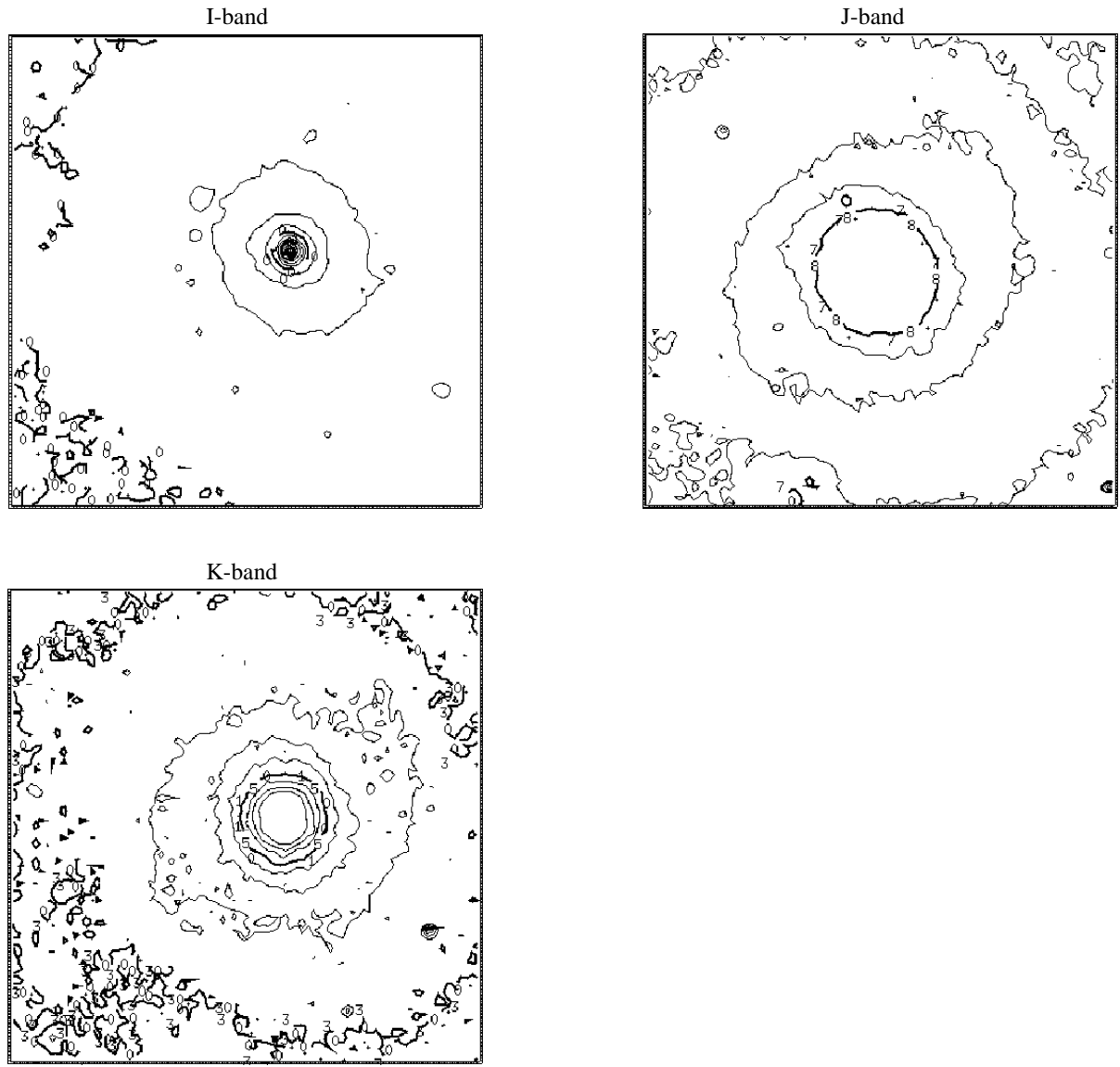}
   \caption{{\em I}, {\em J} and {\em K} band contour plots of Messier 74}
   \label{icontour}
   \end{center}
   \end{figure*}

Figure 3 shows a plot of ellipticity of the contours shown in Figure 2 versus 
minor-axis radius for both the {\em I} band images (circles), the {\em J} 
band image
(triangles) and the {\em K} band image (squares). 
Table 1 shows the same data. Both Figure 3 and Table 1 
show how the ellipticity of the {\em I} band contours increases 
from the centre outwards and then decreases again at larger radii. 
The {\em J} and
{\em K} 
band contours continue to increase, but their position angle changes abruptly
from approximately 45$^{\circ}$ to -55$^{\circ}$ between a radius of 35 and 40
arcsec. This is not seen in the {\em I} 
band, probably due to the increase in dust extinction in this waveband. This
suggests that M74 may contain a {\em bar within a bar}, i.e. a smaller scale
bar which is almost perpendicular to the large scale bar. This is a common
feature often seen in high resolution images of barred galaxies with 
circumnuclear rings of star formation (e.g. Laine et al. 2002).

What we seem to have uncovered here is an oval distortion, rather than a 
strong bar. Is this oval distortion strong enough to create the circumnuclear 
star formation ring seen in M74?

It has been shown in simulations of barred galaxies, that the process of
funneling gas towards the central regions of galaxies by a bar, eventually
leads to the self--destruction of the bar (Friedli \& Pfenniger 1991). 
Bar distortions are useful for driving gas towards an 
ILR, where the gas will then
slowly accumulate. As this process continues, the gas can become 
gravitationally unstable and fall deeper into the potential well (see the
double bar model in Friedli \& Martinet 1993). This accumulation of material
would eventually trigger a burst of star formation, thus leading to a metal
rich, red inner stellar disk. The gas could fall further inwards and create
a central mass concentration. If the mass of the central accretion becomes
significant an extended ILR can be formed leading to dissolution of the bar
(Friedli \& Benz 1993; Friedli 1994). When the bar is dissolved signatures of 
its existence are often seen, e.g. boxy or peanut shaped bulges (e.g. Bureau 
\& Freeman 1999) or 
a metal rich inner stellar disk, as seen in the Sombrero galaxy (Emsellem 
1995; Emsellem et al. 1996). However, these signatures are usually only 
observed in edge--on galaxies. It is therefore difficult to tell if the
oval distortion seen in M74 is part of the progression of the dissolution of 
a bar.

   \begin{figure}
   \begin{center}
   \includegraphics[width=6.5cm,angle=-90]{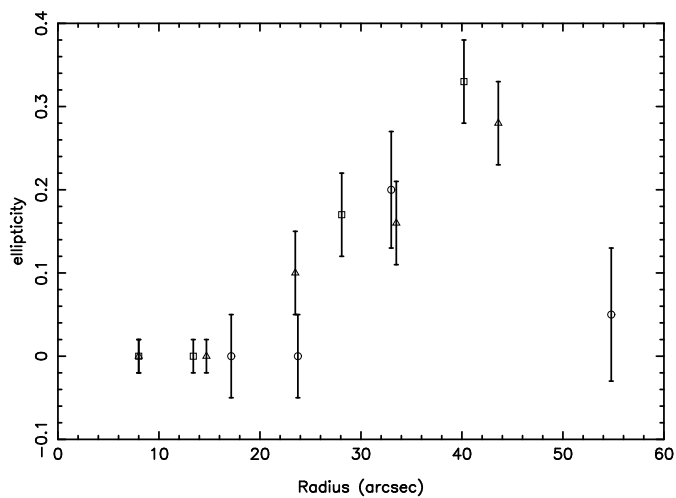}
   \caption{Radius versus ellipticity of the isophotal contours of M74. The 
      circles represent the ellipticities of the isophotal contours in the 
      the {\em I} band image, triangles the {\em J} band image and squares 
      the {\em K} band image.}
   \label{rad_vs_ell}
   \end{center}
   \end{figure}

   \begin{table}
      \begin{center}
      \caption[]{Ellipticity versus radius in M74}
         \label{ellip}
         \begin{tabular}{llll}
            \hline
            \noalign{\smallskip}
            Radius 	& \multicolumn{3}{c}{Ellpticity}	\\
	    (arcsec)	&	{\em I} band & {\em J} band & {\em K} band \\
            \noalign{\smallskip}
            \hline
	    13.4	& --		& 0.00$\pm$0.02 & 0.00$\pm$0.02 \\ 
            17.2	& 0.00$\pm$0.05	& -- & --	\\
	    23.5	& --		& 0.10$\pm$0.05 & -- \\    
            23.8	& 0.00$\pm$0.05 & -- & --	\\
	    28.1	& -- 		& -- & 0.17$\pm$0.05 \\
            33.0	& 0.20$\pm$0.07 & 0.16$\pm$0.05 & -- \\
	    40.2	& -- & -- & 0.33$\pm$0.05	     \\
	    43.6	& -- & 0.28$\pm$0.05	& -- \\
            54.8	& 0.05$\pm$0.08	& -- & --	\\
            \noalign{\smallskip}
            \hline
         \end{tabular}
      \end{center}
   \end{table}

Weliachew et al. (1988) have suggested that even a weak bar, such as the bars
they infer in NGC 6946 and Maffei 2, and the bar observed in M74 here, can 
affect the radial distribution of the gas. Combes \& Gerin (1985) simulated
the behaviour of an ensemble of molecular clouds in a barred galaxy, using an
{\em N}--body barred potential and a collisional scheme for gas clouds. They
used different values for the angular velocity of the bar and showed that, when
the angular velocity of the bar is slow, 
spiral structure develops inside the corotation radius. Angular
momentum is transferred inwards, and particles are trapped in a central orbit
coincident with the ILR. Gaseous material can then accumulate at the ILR, 
forming a circular ring, if the bar angular velocity, 
$\Omega_p \simeq (\Omega - \kappa/2)$, 
where $\Omega$ is the material velocity and
$\kappa$ is the epicyclic frequency. The results of the Combes \& Gerin (1985)
simulations suggests that circumnuclear rings of star formation can exist,
even in the presence of a weak bar potential. Thus, the oval distortion we have
observed in M74 here, may be responsible for the circumnuclear star formation
activity observed by Wakker \& Adler (1995).


\section{Conclusions}

We have presented in this paper, 
$I$, $J$ and $K$ band images 
of the spiral galaxy, M74.
These images highlight 
the presence of a weak oval distortion in the central
regions of the galaxy. We believe that this oval distortion is responsible for
the circumnuclear ring of star formation observed in molecular CO emission
(Wakker \& Adler) and 2.3 $\mu$m CO absorption (James \& Seigar 1999). We have
also argued that even weak oval distortions can be responsible for such
circumnuclear star formation, using the simulations presented by Combes \&
Gerin (1985).


\begin{acknowledgements}
This research is based on data from the ING archive. This research has made 
use of the NASA/IPAC Extragalactic Database (NED) which is operated by the 
Jet Propulsion Laboratory, California Institute of Technology, under contract 
with the National Aeronautics and Space Administration. The author would
like to thank the anonymous referee for comments which greatly improved the
content of this paper.
\end{acknowledgements}

\end{document}